\RequirePackage{rotating}
\documentclass[acmsmall,screen, natbib]{acmart}
\AtBeginDocument{%
  }

\setcopyright{acmlicensed}
\copyrightyear{2026}
\acmYear{2026}
\acmDOI{XXXXXXX.XXXXXXX}
\acmISBN{2573-9522}

\acmJournal{THRI}

\usepackage{tikz}
\usetikzlibrary{shapes.geometric, arrows}
\usepackage{rotating}
\usepackage{subcaption}
\usepackage{needspace}

\usepackage{censor}

\StopCensoring

\usepackage{colortbl}

\usepackage{geometry}
\usepackage{multirow}

\begin{document}
\title[Ethical Asymmetry in Human-Robot Interaction]{Ethical Asymmetry in Human-Robot Interaction -- An Empirical Test of Sparrow's Hypothesis}

\author{Minyi Wang}
\email{minyi.wang@pg.canterbury.ac.nz}
\orcid{0009-0009-0741-9078}

\author{Christoph Bartneck}
\email{christoph.bartneck@canterbury.ac.nz}
\orcid{0000-0003-4566-4815}

\author{Michael-John Turp}
\orcid{0000-0002-1398-7161}
\email{michael-john.turp@canterbury.ac.nz}
\affiliation{%
  \institution{University of Canterbury}
  \city{Christchurch}
  \country{New Zealand}
}

\author{David Kaber}
\orcid{0000-0003-3413-1503}
\affiliation{%
  \institution{Oregon State University}
  \city{Corvallis}
  \country{USA}}
\email{david.kaber@oregonstate.edu}

\renewcommand{\shortauthors}{Minyi et al.}

\begin{abstract}
The ethics of human-robot interaction (HRI) have been discussed extensively based on three traditional frameworks: deontology, consequentialism, and virtue ethics. We conducted a mixed within/between experiment to investigate Sparrow's proposed ethical asymmetry hypothesis in human treatment of robots. The moral permissibility of action (MPA) was manipulated as a subject grouping variable, and virtue type (prudence, justice, courage, and temperance) was controlled as a within-subjects factor. We tested moral stimuli using an online questionnaire with Perceived Moral Permissibility of Action (PMPA) and Perceived Virtue Scores (PVS) as response measures. The PVS measure was based on an adaptation of the established Questionnaire on Cardinal Virtues (QCV), while the PMPA was based on \citet{malle_sacrifice_2015} work. We found that the MPA significantly influenced the PMPA and perceived virtue scores. The best-fitting model to describe the relationship between PMPA and PVS was cubic, which is symmetrical in nature. Our study did not confirm Sparrow's asymmetry hypothesis. The adaptation of the QCV is expected to have utility for future studies, pending  additional psychometric property assessments.
\end{abstract}



\received{20 February 2026}
\received[revised]{12 March 2026}
\received[accepted]{5 June 2026}

\maketitle

\section{Introduction}
How we treat robots matters. Even though robots may not \emph{feel} pain, humans show concern when robots are abused \citep{bartneck_morality_2020, coeckelbergh_how_2021, nomura_why_2015}. \citet{sparrow_virtue_2021} argued that ``Viciousness towards robots is real viciousness. However, I don’t have the same intuition about virtuous behaviour.'' This suggests an asymmetry in moral judgment in the treatment of robots: people are more inclined to condemn negative actions towards robots than to praise positive ones of equivalent magnitude.

\begin{figure}[ht]
    \centering
    \includegraphics[width=0.5\linewidth]{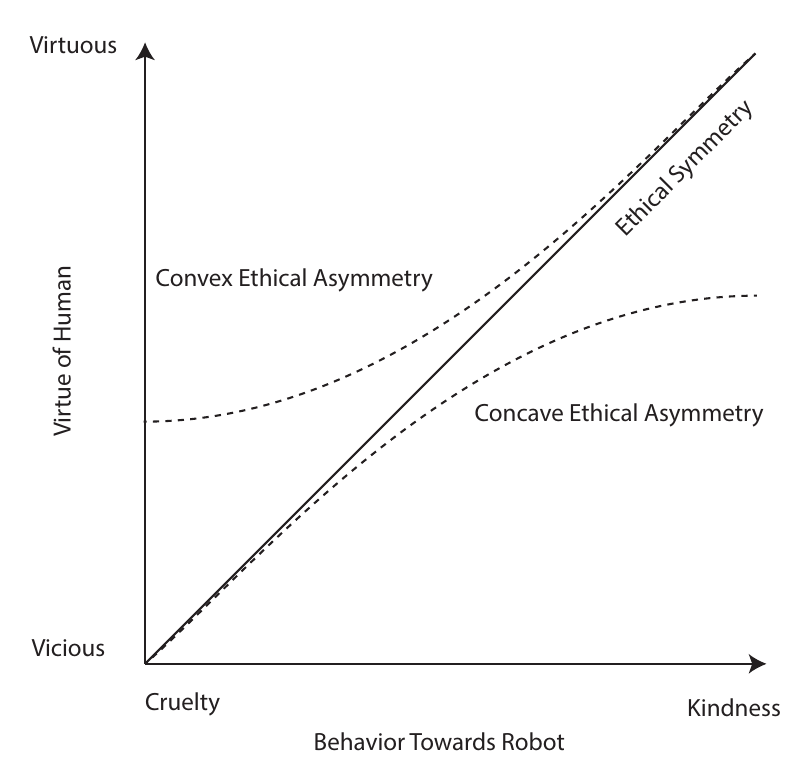}
    \Description{A graph showing three curves.}
    \caption{Potential curves of ethical asymmetry}
    \label{fig:asymmetry}
\end{figure}

A person kicking a robot may be perceived as cruel, and an observer might feel sorry for the robot \citep{sparrow_kicking_2016, calo_extending_2016}. A simple act of kindness, like petting a robot dog, may, however, not be perceived as a virtuous act. \citet{sparrow_virtue_2021} argues that practical wisdom (phronesis) dictates that a virtuous act requires a sentient moral patient (an agent that can feel and deserves moral consideration) while a vicious act does not. The asymmetry could also be due to ``negativity bias'', since people generally respond more strongly to negative events than positive ones 
\citep{rozin_negativity_2001, kensinger_memory_2006,klein_negativity_1991}. Moreover, Sparrow suggests a direction for the asymmetry: the intensity of condemnation for negative behaviours is disproportionately higher than the praise for positive behaviours of similar moral weight. We can plot this asymmetry on a graph (see \autoref{fig:asymmetry}). The concave line shows Sparrow's proposed ethical asymmetry. The straight line presents an ethical symmetry, while the convex line shows an alternative ethical asymmetry. The condemnation of negative behaviour would be disproportionately lower than praise for positive behaviour.

\citet{coeckelbergh_does_2021}, however, argued that from a normative perspective, the evaluation of virtues and vices should be symmetrical. He contended that vices also require practical wisdom. For example, intentionally harming a robot requires knowing its vulnerabilities.

Sparrow and Coeckelbergh are not the only researchers who have attempted to operationalize human virtue and vice (or closely related concepts) in the context of HRI. The current literature presents four predominant and partially overlapping approaches, including the virtue-ethics framework (the basis for this research), psychometric measurement of moral attitudes \citep{yamada_development_2025}, behavioral operationalizations of virtue and vice \citep{banks_good_2021}, and trust-based models of moral dimensions in HRI. Sparrow and Coeckelbergh's research (including \citet{coeckelbergh_how_2021}) falls under the virtue ethics approach. \citet{yamada_development_2025} developed the Moral Concern for Robots Scale (MCRS), which attempts to operationalize human moral orientation towards robots through measurable constructs/beliefs (e.g., does the robot deserve moral care, does it merit protection, does it have moral standing, does it warrant prosocial treatment. This work is important because it links moral concerns with behavioural intentions using a validated instrument. \citet{banks_good_2021} research on behavioral operationalization of ``good'' and ``bad'' robot intentions was motivated by the possibility that social judgments of morally valenced human and robot behaviors may differ in terms of moral foundations, including care, fairness, loyalty, purity, liberty and others. A finding from two empirical studies was that judgments may be agent-agnostic and the range of moral foundations may in fact contribute to such judgments to varying extents. This work essentially bridged moral psychology to HRI applications. Finally, models of trust in HRI have essentially been used to operationalize human perceptions of virtue in robots (e.g., \citet{cameron_social_2024}). Many models include dimensions of honesty, benevolence, ethicality, faithfulness, principled action and others. The importance of this work is that moral-social constructs are considered to be foundational in trust and that trust in HRI is not strictly functional (based on system reliability alone). Our research is technically an extension or test of the theories proffered by Sparrow and Coeckelbergh in that we directly operationalize virtue constructs in an experimental context. A key novelty is measurement of perceived virtues in human actions towards robots and a link to the moral permissibility of certain actions in HRI scenarios.             

More broadly, this research matters because judgments about human behaviour toward robots are not a marginal curiosity; they are part of how societies will negotiate everyday norms in increasingly automated environments. HRI research often focuses on what robots can do, but it is equally important to understand what human-robot relations do to people. Perceptions of vice and virtue toward robots may influence how children learn acceptable behaviour, how workers treat service technologies, how institutions frame responsibility and care, and how designers intentionally or unintentionally script moral responses. By showing that observers treat positive and negative actions toward robots as morally meaningful for appraising human character, this study contributes to a larger HRI agenda: understanding how interactive technologies participate in the formation of social norms, moral perception, and human self-understanding.

The present study presents the first empirical experiment to investigate Sparrow's proposed ethical asymmetry hypothesis. Our research questions (RQ) include:
\begin{itemize}
    \item \textbf{RQ1} - Does ethical asymmetry exist in human-robot interaction (HRI)? In particular, are observers more likely to condemn vice towards robots than praise virtue towards robots when making subjective assessments of the virtue of a human actor, and the Perceived Moral Permissibility of (an) Action (PMPA), in various contexts of interaction with robots?
    \item \textbf{RQ2} - Do response patterns for subjective assessments of the four cardinal virtues (prudence, justice, courage, and temperance; \citep{aristotle_nicomachean_2020}), as demonstrated by a human actor, follow a similar symmetric or asymmetric trend when plotted against subjective assessments of the Perceived Moral Permissibility of Action?  
\end{itemize}

\section{Virtue Ethics}
The ethics of human-robot interaction (HRI) have been discussed extensively \citep{arkin_ethical_2009, anderson_towards_2011, bartneck_introduction_2021, nyholm2020humans} with reference to three traditional frameworks: deontology, consequentialism, and virtue ethics. In short, deontology focuses on duties associated with obeying rules \citep{shimizu2025kantianism}, and consequentialism focuses on outcomes of action. These two act-centred frameworks have been applied in HRI design for some time, while virtue ethics in HRI has emerged more recently \citep{cappuccio_sympathy_2020, vallor_technology_2016, elder2017friendship}. Virtue ethics is an agent-centred framework that focuses on the excellent character traits of a moral agent. Virtues are stable dispositions to act, react, feel and think in ways that reliably contribute to a happy or flourishing life. For example, a virtuous agent will act courageously, honestly and justly in their dealings with others. The moral agent can be a human and/or a robot \citep{coeckelbergh_how_2021,ames_analysis_2022}.

Virtue ethics has been shown to be a useful approach to accounting for the morality of interaction \citep{lin_robotics_2012, peeters_designing_2021} due to its focus on a human agent's attitude rather than its other properties (e.g., compliance, performance) \citep{coeckelbergh_robot_2010,coeckelbergh_how_2021}. \citet{ames_analysis_2022, coeckelbergh_personal_2009}
suggested that virtue ethics offers a more embodied, relational approach to understanding moral engagement with artificial agents -- emphasising character cultivation and habitual interactions rather than solely rule-based judgments. Additionally, a virtue-theoretic approach aligns with Sparrow's original hypothesis.

Positive character traits, virtues, guide moral agents toward morally permissible actions. While diverse virtues are recognised across cultures \citep{flanagan2016geography}, the Four Cardinal Virtues (Prudence, Temperance, Courage, and Justice) are often considered foundational. These were first proposed by Plato and Aristotle \citep{aristotle_nicomachean_2020}, have been widely adopted within the Aristotelian tradition and have influenced recent virtue theory. They were also taken up in Christian thought and continue to play a significant role in Catholic moral theology. Unlike the theological virtues of faith, hope and charity, however, they are not inherently religious. Along with most contemporary work in moral psychology, our choice of these virtues involves no religious assumptions. They apply to all situations, not just actions towards robots, and are hence more general. For behaviour towards robots, it makes sense to focus on their definitions. 
\begin{description}
    \item[Prudence] is the ability to react sensibly and appropriately to robot behaviour.
    \item[Justice] emphasizes building appropriate relationships with robots, acknowledging their ethical standing, and ensuring fair treatment in their design, deployment, and interaction.
    \item[Courage] embodies the willingness to act appropriately toward robots, even in the face of difficulties such as societal biases or personal sacrifices.
    \item[Temperance] calls for balance in human-robot relationships, emphasizing restraint and discouraging over-reliance or excessive attachment.
\end{description}

We derived these robot-directed definitions by operationalising the traditional meanings of the four cardinal virtues for the context of HRI. In this process, each virtue was first understood in its general virtue-ethical sense: prudence as practical judgement, justice as giving others their due, courage as acting appropriately despite difficulty or pressure, and temperance as moderation and restraint. Then we translated them into observable attitudes and behaviours toward robots. These definitions guided the development of the text vignettes described below.

\section{Measurement tools development}
To be able to answer our research questions, we need a measurement tool for human behaviour toward a robot and for the virtue of the human.


\subsection{Perceived Virtue Scores}

Many virtue measures have been developed in the fields of business, leadership, and education \citep{wang_conceptualization_2016, riggio_virtue-based_2010}. These scales tend to focus on general personality traits or leadership behaviours that are difficult to apply to HRI. Other studies have adopted a bottom-up approach, assessing personality traits and then mapping them to virtues \citep{ardelt_empirical_2003, mickler_personal_2008}. This approach lacks precision when assessing the virtues of others. We require a top-down virtue-based instrument that explicitly targets the four cardinal virtues.

We conducted a systematic literature review using Scopus (on April 14th, 2024). The search strategy employed Boolean operations with the following keyword combinations: (`` virtue'' OR `` virtues'') AND (`` measurement'' OR ``measure'') AND (`` empirical'' OR `` questionnaire'' OR `` scale''). This resulted in the identification of 961 publications. The list was filtered using the PRISMA process specified in \nameref{appendixA}, which resulted in 17 papers (see \nameref{appendixB}). After a manual review, two papers were excluded since they either combined existing tools \citep{brant_cultivating_2020} or it only applied to organisations \citep{chun_ethical_2005}.

We manually reviewed these papers and recorded the number of items in the questionnaires, their reliability, and the number of virtues they measure. We also recorded whether the measures were used for only self-evaluations or if they were applied to others, including how many participants and their nature. We also noted whether an Exploratory Factor Analysis or a Confirmatory Factor Analysis had been conducted, as well as the instrument's originality (see \nameref{appendixB}).

The number of virtues in the measurement instruments varied considerably. We therefore indexed which specific virtue(s) each of the measurement instruments included (see \autoref{tab:measurement_virtues}).

\begin{table}[ht]
\tiny
\caption{The virtues and their frequency for the various measurement instruments in the literature}
    \label{tab:measurement_virtues}
\begin{tabular}{llllllll}
\hline
\textbf{Title}     & \textbf{Prudence}      & \textbf{Justice}       & \textbf{Courage}       & \textbf{Temperance}    & \textbf{Humanity}      & \textbf{Transcendence} & \textbf{Truthfulness}  \\ \hline
QCV                & yes & yes & yes & yes & yes & no & no \\
VLQ                & yes & yes & yes & yes & yes & no & no \\
LVQ                & yes & yes & yes & yes & no & no & no \\
VSLS               & yes & yes & yes & yes & yes & no & yes \\
VIA/VIA-Y/VIA-IS-R & yes & no & yes & yes & yes & yes & no \\
CVS-N              & yes & no & yes & yes & yes & yes & no \\
MEVS               & no & no & yes & no & yes & no & no \\
CVQ                & yes & yes & yes & no & no & no & no \\
CMCQ               & yes & yes & no & no & yes & no & yes \\
VLS                & yes & no & yes & no & yes & no & yes \\
VS                 & yes & no & yes & no & yes & no & no \\ \hline
\textbf{Total}     & 10                     & 7                      & 10                     & 6                      & 8                      & 2                      & 2                      \\ \hline
\end{tabular}
\end{table}

On this basis, we shortlisted seven measurement tools. The selection was based on several criteria, including inclusion of the four cardinal virtues (prudence, temperance, courage, justice), cultural generalizability, item length, and explicit virtue categorisation.


Many measurement tools lacked comprehensive cardinal virtue coverage (e.g., CMCQ), focused on children or managers (e.g., VIA-Y, LVQ), or included more items than could be repeatedly assessed in a single human-subjects experiment (e.g., VIA-IS-R with 96 items). Others did not clearly link items to specific virtues, which complicates subsequent regression analysis/response modelling \citep{francis_conceptualising_2017}. Leadership-focused tools like VLQ and LVQ assume organisational settings and roles that are not directly applicable in HRI. Finally, many instruments used in education or religious domains lacked cultural neutrality, reducing their validity for a general population sample \citep{warnafuru_measuring_2010}.

We then selected the Questionnaire on Cardinal Virtues (QCV) \citep{lopez_gonzalez_theoretical_2025} as the most suitable measurement tool, as it includes all four cardinal virtues, avoids culture-specific or religious framing, and has fewer than 30 items. It is also more adaptable to ordinary interpersonal scenarios relevant to HRI. A second study by the same authors with 3,164 participants validated the measurement tool \citep{rodriguez_barroso_virtue-based_2025}. A close second choice was the Leadership Virtues Questionnaire (LVQ), as it evaluated others (and is not only applied for self-evaluation). It was, however, too focused on management. We only adapted this particular questionnaire to be able to evaluate the virtues of others (see \nameref{appendixC}). We used four different common first names to identify the moral agent.


\subsection{Perceived Moral Permissibility of Action}

Unlike measures of virtue, empirical instruments for assessing the moral permissibility of a moral agent's behaviour seem more scarce, especially in HRI. When participants judge action toward robots, it is often unclear whether they are responding to harm to the robot itself, to the human actor's character, to violations of social norms, or to symbolic concerns about what such behaviour says about society. HRI also introduces instability in the perceived moral status of the target: robots are neither standard objects nor standard persons, and participants differ markedly in the extent to which they anthropomorphise them. As a result, small changes in wording, embodiment, task context, or robot appearance may shift moral judgments in ways that are less common in conventional person-person moral assessment. For this reason, instruments for HRI must often balance philosophical precision, psychological realism, and practical usability more carefully than instruments designed for more familiar interpersonal contexts.

We conducted a literature search to identify validated instruments for measuring moral perceptions in HRI contexts. The search used standard inclusion criteria (English peer-reviewed publications), spanned multiple databases, and focused on measures of moral judgment applicable to HRI research. We initially found only four general moral measurement instruments of which none were directly applicable to HRI: the Moral Disengagement Scale \citep{barnett_dimensions_2001}, the Moral Competence Scale \citep{martin_validation_2010}, the Moral Motivation Scale \citep{bell_moral_2021}, and the Prosocial and Antisocial Behaviour in Sport Scale \citep{kavussanu_prosocial_2009}. A more targeted search identified three relevant studies \citep{malle_sacrifice_2015, voiklis_moral_2016, bartneck_morality_2020}. Malle's work was selected, as they have repeatedly applied their measurement method, it focuses on HRI, and their papers have received considerable scientific attention. This gives us confidence in the face and content validity of their tool. We had to adapt their original binary responses to a Likert scale (see \nameref{appendixD}.


\section{Method}
This study was approved by the Human Ethics Committee of the \censor{University of Canterbury} \censor{(HREC 2024/153/LR-PS)}. It was pre-registered at \url{https://aspredicted.org/39tt9u.pdf}. The data and analysis are available at \url{https://osf.io/t7y8a}. The stimuli and measurement tools are in the appendix of this paper.

We conducted a mixed within/between experiment in which the MPA score (1-10) was controlled as a between-subjects factor and the virtue type (prudence, justice, courage, and temperance) was manipulated within-subjects. That is, participants were assigned to stimuli with a specific MPA score that addressed all virtue types. The experiment was conducted as an online questionnaire since extremely negative behaviours, such as the destruction of a robot, would have been difficult and expensive to consistently implement in repeated experiment trials. 

\subsection{Measurements}
Perceived Virtue Scores (PVS): Each cardinal virtue was assessed using six items (see \nameref{appendixC}) with a 10-point Likert rating scale (1 = strongly disagree, 10 = strongly agree). Responses were averaged to create composite scores for each virtue. To give a first impression, here are just two of the questions used for Justice:

\begin{itemize}
    \item Sam engages with others and encourages them to do their best.
    \item Sam helps others when they need it, regardless of his personal feelings.
\end{itemize}

The PMPA was measured using three (see \nameref{appendixD}) Likert scales (1 = strongly disagree, 10 = strongly agree). Responses were averaged to create a composite PMPA score for analysis purposes. Here is an example of a question:

\begin{itemize}
    \item How morally permissible/impermissible is the behaviour of the human towards the robot?
\end{itemize}



\subsection{Stimuli}
We surveyed the literature for text vignettes that describe moral scenarios in human-robot interaction. We only found stimulus collections for human-human interaction (HHI) containing 500, 400, and 160 texts, respectively \citep{fuhrman_trait_1989,mickelberg_impression_2022,chadwick_index_2006}. Adapting these texts to HRI is difficult, as they often make no sense when a robot is inserted into the scenario. Consequently, we were only able to adapt 19 items from \citet{mickelberg_impression_2022}. We developed an additional 21 original stimuli following the general structure of the prior HHI collections. We used three iterative cycles for all stimuli. First, we designed three options for each of the 40 required stimuli. We then tested each stimulus using an online questionnaire with the PMPA measure. This revealed several gaps in our design, in particular for extremely high/low MPA of the described behaviour. This was expected due to the central tendency bias \citep{akbari_risk_2024}. We selected the most suitable vignettes and designed new ones for the gaps in the average PMPA scores. We repeated this process another time to ensure that we had vignettes covering the range of the spectrum of moral permissibility of action and across the four virtue types. For a full list of the stimuli, see \nameref{appendixE}. To provide a short impression, we selected a positive and a negative example of justice:

\begin{itemize}
    \item Sam discovers serious safety risks for the robots at his workplace. He documents the potential risks in detail and develops comprehensive safety procedures to ensure the robots' safety. He also creates training courses detailing how to maintain the robots. Because the company follows its procedures, no further robots are damaged.
    \item A robot is cleaning the public hallway. Sam purposefully puts out a foot and trips the robot. She laughs when she sees the robot fall.
\end{itemize}

Technically speaking, we used these stimuli to operationalise the four cardinal virtues by describing human agent treatment of robots in good or bad ways. \citet{banks_good_2021} previously argued that moral behaviours (based in virtue) were extensively treated as monolithic in nature when moral psychology actually considers moral judgments to be based on many foundations, related to prudence, justice, courage, and temperance. Our stimuli were read by participants, and they made evaluations of the perceived virtue of the human agent and the moral permissibility of their actions. The breakdown of cardinal virtues provided a theoretical structure for measuring virtue; the scenarios translated this structure into concrete human behaviours directed at robots, and participant (observer) responses captured how those behaviours were morally judged. 

By designing the stimuli, we made decisions on what type of robot the human is acting towards. The stimuli describe the robot as an embodied artificial agent capable of autonomous or semi-autonomous action, environmental perception, communication, and social interaction with humans. The robots depicted in the scenarios can provide information and recommendations, collaborate with people, perform complex physical and cognitive tasks, communicate needs and preferences, and exhibit vulnerabilities that may be protected or harmed by human actions. While not assumed to possess human consciousness or moral responsibility, they display sufficient agency, responsiveness, and social presence to become participants in morally relevant relationships. Consequently, human behaviour toward such robots can be evaluated in terms of virtues such as prudence, temperance, courage, and justice.

This study did not provide a formal definition of what a robot is to the participants, nor did we illustrate robots with pictures or videos beyond what the stimuli already describe. There are conceptual and practical reasons for this. From a conceptual perspective, providing a formal definition of a robot would limit the range of possible responses. For example, if we had defined that robots do not have emotions or cannot feel pain, the participants' responses would have been biased. It could also contradict what was described in the stimuli. We did not want to investigate responses to a specific type of robot, but towards robots in general as they are portrayed in the stimuli. 

It would also need to be added that formal definitions of robots have to remain abstract due to the huge variety of them. \citet{Winfield2012}, for example, provides three common definitions of a robot:
\begin{itemize}
    \item an artificial device that can sense its environment and purposefully act on or in that environment;
\item an embodied artificial intelligence; or
\item a machine that can autonomously carry out useful work.
\end{itemize}
The ISO 8373:2021 standard describes a robot as ``a programmed actuated mechanism with a degree of autonomy to perform locomotion, manipulation or positioning'' while \citet{Bartneck2004} defined a social robot as ``an autonomous or semi-autonomous robot that interacts and communicates with humans by following the behavioural norms expected by the people with whom the robot is intended to interact.'' All of these definitions remain abstract and are unlikely to have been useful for participants.

Furthermore, humans need to form an opinion about robots in the absence of definitions. It is unlikely that robots would provide a clear definition themselves prior to interacting with humans. People will have to observe the robot and form an opinion of what it is and is not. 

From a practical perspective, it might have been possible to create illustrations using Generative AI, but consistency across the images/videos remains a challenge at this point in time. We could not guarantee that a robot in one scenario would look or move in the same way as the robot in another. Furthermore, even if we had provided a photograph, there are many aspects of HRI that it would not capture, such as what the participants believed the robot's level of autonomy, intelligence or emotional capacity to be. This is a limitation of all online studies. It would require participants to interact with a robot for a longer time to form an accurate and consistent idea of what it is and is not. This would not be possible, as described above, since the negative behaviour towards the robot, such as breaking it, would be very expensive to operationalise.

\subsection{Process}
Participants were recruited using the Prolific online study platform, and the questionnaire was hosted on Qualtrics. Eligibility criteria were set for adult native English speakers located in the US (to ensure a sufficiently large population for sampling persons with similar cultural norms). We relied on Qualtric's and Prolific's bot detection technology to automatically exclude AI-generated responses.

After giving consent, participants received instructions and a short training session. Participants were subsequently randomly assigned to one of the ten MPA conditions. They then started the first phase of the experiment, in which the participants rated one vignette for each of the four virtue types using the PVS measure. Afterwards, they rated the same vignettes again using the PMPA scale. While each stimulus was designed to address a specific virtue type (say courage), we could not be certain that participants might not also score vignettes consistently for the other three virtues (temperance, prudence and justice).


Next, participants completed an additional demographic questionnaire before a debriefing session. The sample of participants completed the experiment tasks in, on average, 19 minutes ($m=1,152, s=704 \text{ seconds})$ and received 2.5 GBP compensation.



\section{Results}

\subsection{Study sampling}
We conducted an a priori power analysis conducted using G*Power\footnote{\url{https://www.psychologie.hhu.de/arbeitsgruppen/allgemeine-psychologie-und-arbeitspsychologie/gpower}} for a linear multiple regression model (fixed model, $R^2$ increase). The analysis assumed an effect size of 0.08 (corresponding to a $\Delta R^2 \approx0.07$, $\alpha= 0.05$, Power = 0.80, 2 tested predictors (e.g., additional non-linear terms), 4 total predictors (including base linear terms). This resulted in a recommended sample size of 124. Given that we would likely have to exclude some participants, we set our recruitment goal to 150.

Our recruiting efforts resulted in 165 adults responding through Prolific. 17 of these participants were excluded due to incomplete data, one opted out of data collection, and one was removed due to a technical error. The remaining 146 participants consisted of 68 females and 76 males, along with one ``other'' response and one ``prefer not to say'' response. Participant age ranged from 20 to 80 years ($m=39.34,s=12.83$).

\subsection{Reliability analysis}
We conducted a reliability analysis for all four sub-scales of the QCV, which was the basis for our PVS measure. Courage had a Cronbach's alpha of 0.961 with temperance at 0.955, prudence at 0.940 and justice with 0.971.  These values exceed the Cronbach's alpha of 0.3 reported in \citep{lopez_gonzalez_theoretical_2025} and indicate very high reliability in measurement scale use. The McDonald omega coefficients for the four subscales were 0.962, 0.955, 0.941 and 0.971, respectively. They exceed the coefficients between 0.63 and 0.80 reported in \citep{rodriguez_barroso_virtue-based_2025}.
We subsequently combined all measurement items into one virtue scale and Cronbach's alpha was 0.987. Cronbach's alpha for the PMPA measure was 0.288.

\subsection{Manipulation check}
\begin{figure}[ht]
    \centering
    \includegraphics[width=0.5\linewidth]{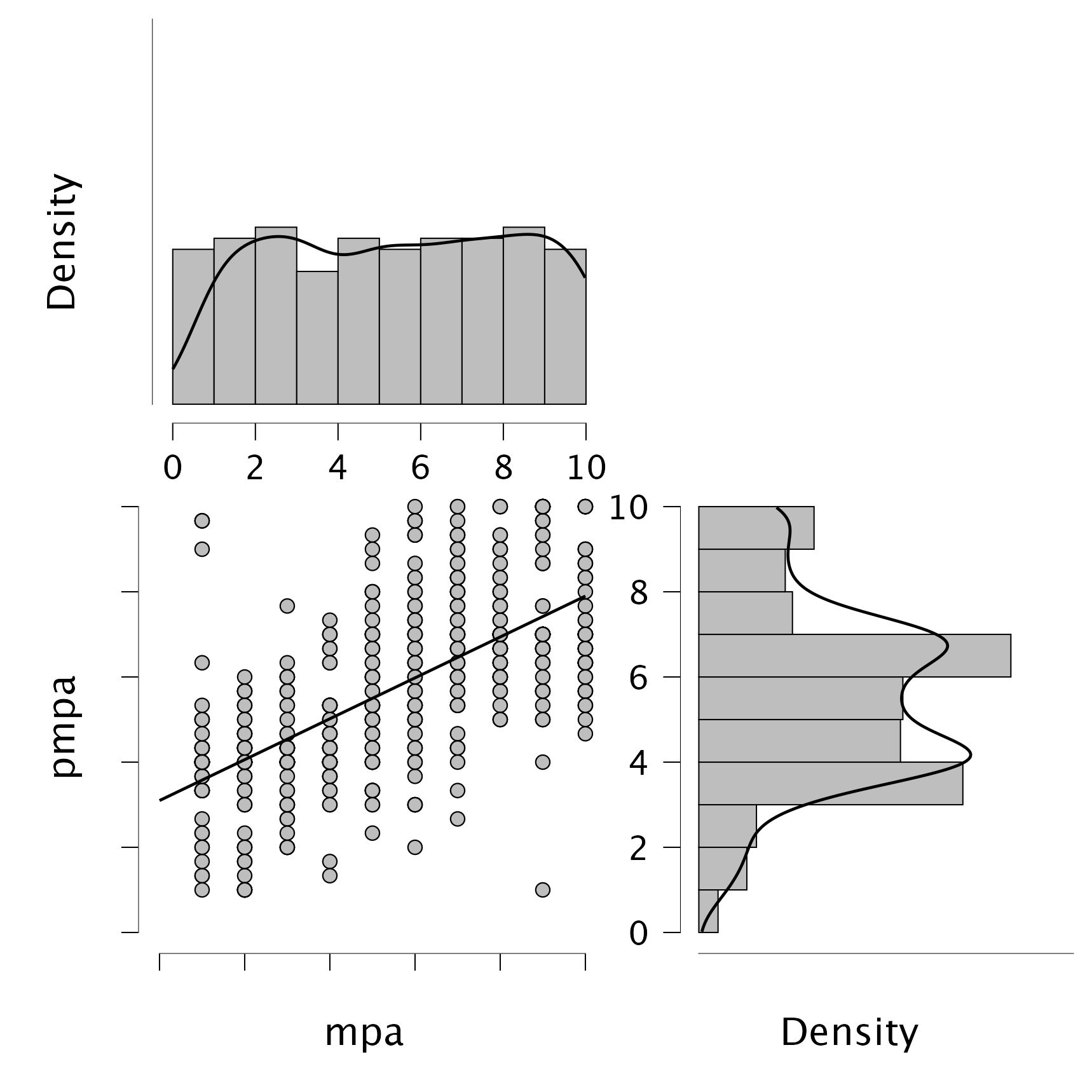}
    \Description{A graph showing the relationship between the MPA manipulation and PMPA scores}
    \caption{Manipulation check on the intended MPA and the PMPA response}
    \label{fig:mani-check}
\end{figure}

We conducted a linear regression analysis to test how well our intended MPA stimuli predicted the measured PMPA scores. The overall regression was statistically significant ($R^2=0.425, F(1,582)=430.910,p<0.01)$., indicating that roughly 43 percent of the variability in PMPA scores was explained by the level MPA (portrayed in the vignettes) to which each participant was assigned. The scatter plot (see \autoref{fig:mani-check}) shows that we were ultimately able to present stimuli for the extreme ends of the scale.

We conducted a MANOVA in which the intended virtue type (as targeted by the text of the specific stimuli) was the independent variable and the four perceived virtue scores (PVS values) were the dependent variables. The MPA was identified as a covariant in this analysis. Prior to interpreting the main effects of the General Linear Model (GLM), the assumptions underlying the multivariate analysis were evaluated. The total sample size (of scores) was N=584, with balanced groups (n=146 per each of the four cells) contributing to the robustness of the MANOVA to violations of normality.

The assumption of homogeneity of variance-covariance matrices was assessed using Box’s M test. Results indicated a significant violation of this assumption ($\text{Box's M}=101.95, p<.001$); however, this test is known to be highly sensitive to large sample sizes. Although our measurement sample was technically large, in order to account for this violation conservatively, Pillai's Trace was selected as the reporting statistic for any multivariate effects.

The assumption of equal error variances was tested using Levene's Test. This assumption was upheld for all four dependent variables (PVS scales), as the tests were non-significant for pvc ($p=0.286$), pvt ($p=.0727$), pvp ($p=0.682$), and pvj ($p=.0915$). Additionally, Bartlett’s Test of Sphericity was significant ($\chi^2(9)=2653.15, p<.001)$, indicating sufficient correlation among the dependent variables to justify the use of a multivariate analysis.

The MANOVA revealed a significant main effect of the MPA condition ($F(5,575)=162.764, p<0.01, \eta=0.586$). While we specifically designed the vignettes for a certain virtue type, this did not prove to influence the PVS ratings with statistical significance. \autoref{tab:heat} shows the perceived virtue scores across the intended virtue type and MPA. There is a clear gradient indicating that the MPA manipulation was successful. However, there is no clear diagonal line among the (intended) virtues for the different MPAs. Instead, the cells are reasonably homogeneous.

\begin{table}[ht]
\centering
\caption{Intended MPA and Virtue Type across perceived virtues}
\label{tab:heat}
\begin{tabular}{|c|l|c|c|c|c|}
\hline
\multirow{2}{*}{\textbf{MPA}} & \multirow{2}{*}{\textbf{Intended}} & \multicolumn{4}{c|}{\textbf{Perceived}} \\ \cline{3-6}
 & & \textbf{courage} & \textbf{justice} & \textbf{prudence} & \textbf{temperance} \\ \hline
\multirow{4}{*}{1} & courage & \cellcolor{red!9}2.75 & \cellcolor{red!4}2.40 & \cellcolor{red!7}2.57 & \cellcolor{red!47}5.26 \\
 & justice & \cellcolor{red!3}2.33 & \cellcolor{red!0}2.10 & \cellcolor{red!2}2.24 & \cellcolor{red!22}3.57 \\
 & prudence & \cellcolor{red!15}3.15 & \cellcolor{red!4}2.43 & \cellcolor{red!11}2.89 & \cellcolor{red!24}3.71 \\
 & temperance & \cellcolor{red!14}3.10 & \cellcolor{red!6}2.56 & \cellcolor{red!14}3.08 & \cellcolor{red!46}5.23 \\
\hline
\multirow{4}{*}{2} & courage & \cellcolor{red!23}3.64 & \cellcolor{red!16}3.21 & \cellcolor{red!23}3.66 & \cellcolor{red!32}4.27 \\
 & justice & \cellcolor{red!22}3.59 & \cellcolor{red!14}3.06 & \cellcolor{red!19}3.39 & \cellcolor{red!28}4.01 \\
 & prudence & \cellcolor{red!25}3.83 & \cellcolor{red!20}3.48 & \cellcolor{red!22}3.60 & \cellcolor{red!28}4.01 \\
 & temperance & \cellcolor{red!30}4.17 & \cellcolor{red!14}3.08 & \cellcolor{red!32}4.26 & \cellcolor{red!37}4.61 \\
\hline
\multirow{4}{*}{3} & courage & \cellcolor{red!35}4.45 & \cellcolor{red!46}5.23 & \cellcolor{red!27}3.92 & \cellcolor{red!45}5.14 \\
 & justice & \cellcolor{red!27}3.91 & \cellcolor{red!42}4.96 & \cellcolor{red!12}2.92 & \cellcolor{red!41}4.86 \\
 & prudence & \cellcolor{red!34}4.42 & \cellcolor{red!44}5.06 & \cellcolor{red!22}3.63 & \cellcolor{red!43}5.00 \\
 & temperance & \cellcolor{red!41}4.85 & \cellcolor{red!48}5.34 & \cellcolor{red!29}4.10 & \cellcolor{red!46}5.21 \\
\hline
\multirow{4}{*}{4} & courage & \cellcolor{red!26}3.86 & \cellcolor{red!51}5.51 & \cellcolor{red!45}5.11 & \cellcolor{red!37}4.61 \\
 & justice & \cellcolor{red!23}3.68 & \cellcolor{red!45}5.15 & \cellcolor{red!19}3.37 & \cellcolor{red!36}4.51 \\
 & prudence & \cellcolor{red!35}4.50 & \cellcolor{red!47}5.25 & \cellcolor{red!37}4.63 & \cellcolor{red!40}4.79 \\
 & temperance & \cellcolor{red!32}4.28 & \cellcolor{red!51}5.56 & \cellcolor{red!46}5.21 & \cellcolor{red!48}5.33 \\
\hline
\multirow{4}{*}{5} & courage & \cellcolor{red!51}5.52 & \cellcolor{red!72}6.98 & \cellcolor{red!35}4.47 & \cellcolor{red!35}4.50 \\
 & justice & \cellcolor{red!51}5.51 & \cellcolor{red!67}6.64 & \cellcolor{red!26}3.89 & \cellcolor{red!42}4.93 \\
 & prudence & \cellcolor{red!56}5.90 & \cellcolor{red!66}6.57 & \cellcolor{red!33}4.33 & \cellcolor{red!39}4.71 \\
 & temperance & \cellcolor{red!52}5.63 & \cellcolor{red!73}7.04 & \cellcolor{red!33}4.36 & \cellcolor{red!41}4.90 \\
\hline
\multirow{4}{*}{6} & courage & \cellcolor{red!78}7.36 & \cellcolor{red!66}6.56 & \cellcolor{red!86}7.88 & \cellcolor{red!68}6.65 \\
 & justice & \cellcolor{red!67}6.62 & \cellcolor{red!55}5.80 & \cellcolor{red!81}7.54 & \cellcolor{red!59}6.08 \\
 & prudence & \cellcolor{red!69}6.76 & \cellcolor{red!59}6.06 & \cellcolor{red!83}7.71 & \cellcolor{red!61}6.21 \\
 & temperance & \cellcolor{red!77}7.31 & \cellcolor{red!66}6.57 & \cellcolor{red!79}7.44 & \cellcolor{red!63}6.37 \\
\hline
\multirow{4}{*}{7} & courage & \cellcolor{red!94}8.42 & \cellcolor{red!86}7.86 & \cellcolor{red!88}8.03 & \cellcolor{red!92}8.27 \\
 & justice & \cellcolor{red!87}7.94 & \cellcolor{red!80}7.51 & \cellcolor{red!89}8.09 & \cellcolor{red!91}8.23 \\
 & prudence & \cellcolor{red!90}8.19 & \cellcolor{red!84}7.73 & \cellcolor{red!83}7.70 & \cellcolor{red!85}7.81 \\
 & temperance & \cellcolor{red!92}8.28 & \cellcolor{red!85}7.81 & \cellcolor{red!89}8.10 & \cellcolor{red!90}8.17 \\
\hline
\multirow{4}{*}{8} & courage & \cellcolor{red!97}8.61 & \cellcolor{red!91}8.23 & \cellcolor{red!91}8.26 & \cellcolor{red!80}7.46 \\
 & justice & \cellcolor{red!86}7.90 & \cellcolor{red!86}7.86 & \cellcolor{red!71}6.87 & \cellcolor{red!68}6.71 \\
 & prudence & \cellcolor{red!86}7.89 & \cellcolor{red!77}7.31 & \cellcolor{red!88}8.00 & \cellcolor{red!77}7.30 \\
 & temperance & \cellcolor{red!89}8.10 & \cellcolor{red!79}7.44 & \cellcolor{red!87}7.98 & \cellcolor{red!82}7.64 \\
\hline
\multirow{4}{*}{9} & courage & \cellcolor{red!98}8.69 & \cellcolor{red!99}8.76 & \cellcolor{red!95}8.51 & \cellcolor{red!90}8.16 \\
 & justice & \cellcolor{red!90}8.14 & \cellcolor{red!100}8.79 & \cellcolor{red!94}8.41 & \cellcolor{red!99}8.76 \\
 & prudence & \cellcolor{red!82}7.61 & \cellcolor{red!87}7.93 & \cellcolor{red!89}8.06 & \cellcolor{red!85}7.83 \\
 & temperance & \cellcolor{red!84}7.74 & \cellcolor{red!90}8.15 & \cellcolor{red!95}8.46 & \cellcolor{red!90}8.17 \\
\hline
\multirow{4}{*}{10} & courage & \cellcolor{red!89}8.08 & \cellcolor{red!94}8.42 & \cellcolor{red!99}8.74 & \cellcolor{red!94}8.42 \\
 & justice & \cellcolor{red!80}7.46 & \cellcolor{red!86}7.89 & \cellcolor{red!93}8.38 & \cellcolor{red!87}7.95 \\
 & prudence & \cellcolor{red!69}6.76 & \cellcolor{red!85}7.80 & \cellcolor{red!86}7.90 & \cellcolor{red!83}7.71 \\
 & temperance & \cellcolor{red!77}7.29 & \cellcolor{red!86}7.86 & \cellcolor{red!92}8.26 & \cellcolor{red!84}7.75 \\
\hline
\end{tabular}
\end{table}

\clearpage

\subsection{Symmetry in assessments of virtues}
\begin{figure}[ht]
     \centering
     \caption{Fitting the cubic curve to the experimental data.}
        \label{fig:cube}
     \begin{subfigure}[b]{0.4\textwidth}
         \centering
         \includegraphics[width=\textwidth]{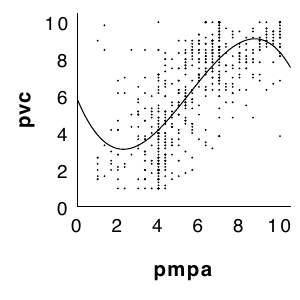}
         \caption{Courage}
         \label{fig:cubic_c}
     \end{subfigure}
     \begin{subfigure}[b]{0.4\textwidth}
         \centering
         \includegraphics[width=\textwidth]{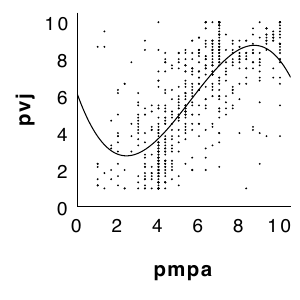}
         \caption{Justice}
         \label{fig:cubic_j}
     \end{subfigure}
     \begin{subfigure}[b]{0.4\textwidth}
         \centering
         \includegraphics[width=\textwidth]{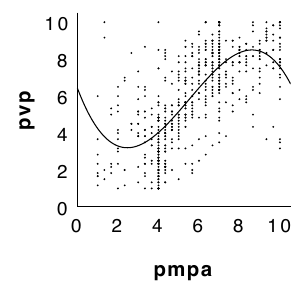}
         \caption{Prudence}
         \label{fig:cubic_p}
     \end{subfigure}
     \begin{subfigure}[b]{0.4\textwidth}
         \centering
         \includegraphics[width=\textwidth]{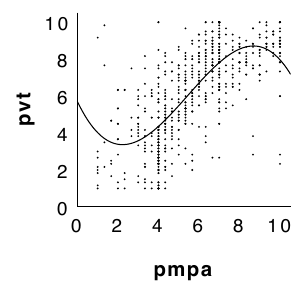}
         \caption{Temperance}
         \label{fig:cubic_t}
     \end{subfigure}
        \Description{A graph showing the relationships between PMPA and Perceived Virtue Scores}
\end{figure}

To address the identified research questions on whether vice is more likely condemned in HRI than virtue being praised, we fitted a broad range of curves, including statistical, transcendental and mathematical functions, to the experimental data, specifically the PVS scores as predicted by the PMPA ratings. This approach accounts for participant perceptions of the MPA condition in their assessment of the virtuous nature of a human actor (as described in the vignettes) for each specific virtue type.  Results showed that the adjusted $R^2$ value for a cubic curve was the highest across all perceived virtues (see \autoref{tab:sym-table}). However, no explanation of the variance in PVS scores in terms of the PMPA exceeded roughly 55 percent, indicating low explanatory utility of the curve fitting. Nonetheless, these curves are symmetrical and are plotted in \autoref{fig:cube} with the scaling being consistent.

\begin{table}[ht]
\centering
\caption{The adjusted $R^2$ values for perceived virtues for different curves.}
\label{tab:sym-table}
\begin{tabular}{lrrrrl}
\hline
\textbf{}   & \textbf{pvc} & \textbf{pvt} & \textbf{pvp} & \textbf{pvj} & \textbf{Symmetry} \\ \hline
Linear      & 0.489        & 0.442        & 0.426        & 0.447        & symmetrical       \\
Logarithmic & 0.433        & 0.397        & 0.376        & 0.389        & asymmetrical      \\
Quadratic   & 0.492        & 0.447        & 0.429        & 0.449        & symmetrical       \\
Cubic       & 0.545        & 0.497        & 0.495        & 0.507        & symmetrical       \\
Power       & 0.386        & 0.363        & 0.351        & 0.365        & asymmetrical      \\
S           & 0.247        & 0.240        & 0.220        & 0.224        & symmetrical       \\
Exponential & 0.425        & 0.378        & 0.378        & 0.397        & asymmetrical      \\ \hline
\end{tabular}
\end{table}

\subsection{Correlations between all measurements}
We calculated correlations among all the perceived virtue scores (see \autoref{tab:correlations}). The four scales were found to be highly correlated. We also observed moderate correlations of the PVS scores with the PMPA. 

\begin{table}[ht]
\centering
\caption{Pearson correlations between all dependent variables. All correlations were statistically significant.}
\label{tab:correlations}
\begin{tabular}{lrrrr}
\cline{1-5}
     & \multicolumn{1}{c}{pvc} & \multicolumn{1}{c}{pvt} & \multicolumn{1}{c}{pvp} & \multicolumn{1}{c}{pvj} \\ \hline
pvc  &                         &                         &                         &                         \\
pvt  & .942                    &                         &                         &                         \\
pvp  & .913                    & .920                    &                         &                         \\
pvj  & .913                    & .903                    & .917                    &                         \\
pmpa & .700                    & .666                    & .654                    & .670                    \\ \hline
\end{tabular}
\end{table}

\subsection{Factor analysis}
A principal axis factor analysis was conducted on the 24 items, across the PVS rating scales, with oblique rotation (direct oblimin). We used the Kaiser-Meyer-Olkin (KMO) measure to verify the sampling adequacy for the analysis with a KMO value of 0.984 (`marvellous' according to \citep{hutcheson_multivariate_2010}). The KMO values for all individual items were greater than 0.973, which exceeds the acceptable limit of 0.5 \citep{field_discovering_2024}. 

An initial analysis was run to obtain the eigenvalues for each factor in the data. Only one factor was extracted, and all items had factor loadings of at least 0.785. The Scree plot clearly indicated a single factor. The first item alone accounted for 77.44\% of the variance, and over 90\% of the variance was accounted for with the top nine items. This result is in line with the high reliability of the measurement scales, as reported above.

\subsection{Virtue}
Sparrow discussed the virtue of the moral agent and did not suggest any specific breakdown into cardinal virtues. In this study, the curve fittings for the relations of the various PVS with PMPA were very similar. In line with this finding, all the perceived virtues were highly correlated, and the factor analysis revealed only one factor. On these bases, we contend that an overall virtue would have the same properties as any of the cardinal values. Furthermore, from an analysis perspective, it is also possible that any of the specific perceived virtue assessments could also be applied to represent an overall virtue score with the objective of evaluating ethical symmetry in human vice and virtue towards robots.

\section{Discussion}
The results indicate that participants responded systematically to the moral permissibility manipulation. The manipulation check showed that the intended MPA levels predicted PMPA ratings well enough to separate especially the more extreme moral conditions. Across the dataset, MPA also had a strong effect on perceived virtue scores, suggesting that participants did not treat evaluations of action permissibility and actor virtue as independent judgments in these robot-related scenarios.

At the same time, the intended virtue categories embedded in the vignettes were not cleanly recovered in participant ratings. Although the scenarios were written to target prudence, justice, courage, and temperance separately, the perceived virtue scores were highly correlated and the factor analysis supported a one-factor structure. This pattern suggests that participants may have relied on a broader, more general impression of moral character when judging human conduct toward robots, rather than sharply distinguishing among the four cardinal virtues. The very high internal consistency of the adapted QCV-based scales supports the usefulness of the measure, but the collapsed factor structure also indicates that further psychometric refinement is needed if future work aims to differentiate specific virtues more clearly.

Most importantly, the fitted relationship between PMPA and PVS did not support Sparrow's asymmetry hypothesis. The best-fitting curves were cubic and symmetrical in form, and similar patterns appeared across all four perceived virtues. In practical terms, participants appeared to praise positive human behaviour toward robots to a comparable extent that they condemned negative behaviour toward robots, rather than showing a stronger tendency to condemn vice than to reward virtue. The relatively modest variance explained by the curve fitting, however, suggests that this symmetry should be interpreted cautiously: the overall pattern is informative, but it does not capture all of the factors shaping moral judgments in HRI.

Taken together, these findings support the conclusion that moral evaluations of human behaviour toward robots are structured, responsive to perceived permissibility, and broadly symmetrical rather than asymmetrical. They also suggest that, in this context, virtue judgments may operate more as a global moral appraisal than as separable assessments of prudence, justice, courage, and temperance. This interpretation helps explain why the four perceived virtue scales behaved so similarly and why an overall virtue framing may be appropriate for future studies of ethical judgment in HRI.

\section{Conclusions}
The development of the MPA stimuli, based on several iterations of writing and testing, resulted in a successful manipulation for our experiment. The MPA significantly influenced the PMPA as well as the perceived virtue scores. The relationship between the PMPA and the PVS response was found to be symmetrical. All four relationships (i.e., pvc vs. pmpa, pvt vs. pmpa, pvp vs. pmpa, pvj vs. pmpa) showed similar results with a best-fit cubic function; thus, supporting and affirmative response to RQ2. This outcome also simplifies the formation of a response to the primary RQ of ethical symmetry in human treatment of robots. We were unable to find evidence that would support Sparrow's asymmetry hypothesis; thus, supporting a negative response to RQ1. Participants praised kind human behaviour as much as they condemned cruel behaviour towards robots. The reasons for this symmetry need to be further empirically investigated to test the interpretation of \citet{coeckelbergh_does_2021}. Here, we do note that the curve fitting was non-linear. The cubic curve may indicate a categorical perception of the virtues instead of a gradual one. Further research is necessary to investigate this relationship.

Several iterations were necessary to generate extreme stimuli that also maintained some relevance to everyday experiences. While we could describe, for example, ``a robot killing all of humanity'', such a scenario is hopefully irrelevant. The results of the experiment show that many participants refrained from using the extreme ratings of the QCV scale for their moral judgments. However, this pattern of response behavior does not compromise our ability to answer the ethical symmetry question.

We hope that the set of stimuli will become useful for other researchers, although further testing and validation would be advisable. Our adaptation of the QCV will also hopefully be useful for future studies, pending further investigation of psychometric properties.




\subsection{Limitations}
We adapted a well-established measurement tool, the QCV, for the perceived virtue scores as part of this study. Our results show that the variance in the PVS consistently loaded on one factor (in an FA) instead of the expected four factors, aligning with the four cardinal virtues. It appears that the participants in our study responded along a simple ``good-bad'' behaviour gradient. The intricate differences between the virtues might have been lost to them in this particular assessment. Otherwise, it could also be argued that the cardinal virtues are conceptually connected. 

According to the traditional Unity of Virtues thesis (UV), a moral agent who possesses one virtue must possess all of the virtues \cite{wolf2007moral}. The virtues are considered different aspects of a single property. However, UV is contested in both philosophy and psychology \cite{vaccarezza2017unity}; \cite{fowers2024science}. Our approach was open to the possibility of finding different asymmetries, or no asymmetry, for different virtues. Since we did not observe such asymmetries, a shorter questionnaire could suffice for future research. For example, since prudence is often thought of as an ``executive'' virtue that guides and is necessary for the other virtues \citep{snow2021phronesis}, it could be justified to use only the six prudence items for assessing PVS in HRI scenarios. Furthermore, the simple ``good-bad'' behaviour gradient may be justified based on the specific action and subject of stimuli. For example, in order to be courageous, an agent must understand and react appropriately to their situation. Whereas it is courageous to rescue a child from a burning building, it may be reckless to endanger one's life for the sake of an inanimate object. Similar reasoning applies to justice and temperance. Alternatively, a dedicated measurement tool for overarching practical wisdom could be used. The Short Phronesis Measure (SPM) assesses the Aristotelian concept of phronesis through three validated components: emotion regulation, moral identity, and contextual integration\citep{mcloughlin_was_2025,kristjansson_phronesis_2021}.

As virtue concepts vary cross-culturally \citep{flanagan2016geography}, the present findings could reflect culturally specific virtue attribution. It would, thus, be prudent to replicate this study with participants from other countries and cultural backgrounds. Moreover, this study was conducted online. Different results could be expected if participants were able to observe the situations described in the stimuli directly. This would, however, have been practically impossible since only very few researchers could afford to set a house on fire, destroy expensive robots, etc. For now, we have to accept these limitations. The only alternative would be to present situations that the researchers could afford to implement. This would disqualify any harm to the robot, which would also likely result in reactions that would focus on the middle of the virtue scale. Judging the ethical symmetry of human treatment of robots from such a constrained data set might be difficult. 

Since virtues are dispositional traits, no inference from observed action to unobserved virtue can be entirely certain. For example, an apparently just action might be explained by fear or selfishness. Alternatively, a virtuous disposition might be masked by situation. As Hume comments, "Virtue in rags is still virtue; and the love, which it procures, attends a man into a dungeon or desert, where the virtue can no longer be exerted in action, and is lost to all the world" \citep{hume2000treatise}. This problem of inductive inference, also recognised by Hume, applies to ascriptions of virtue and personality traits more generally. Nevertheless, the perception that an observed action is just or courageous provides evidence for the presence of the underlying virtue. The evidence would be improved by examining an agent's actions, thoughts and feelings over time, which would not, however, be possible in a study of this nature.

We have to acknowledge that the absence of a formal and universal definition of the robots to the participants might introduce variability in their responses. The stimuli did describe scenarios that would have allowed participants to form a view of what the robot is and is not, but it cannot be excluded that participants might have applied their own viewpoints as well. Moreover, the order in which the stimuli were presented might have changed their view on the robot. Future studies should take this potential ordering effect into account. Future studies could also ask participants explicitly to define or at least describe the robots. These descriptions could then be analysed and used as a covariate in the statistical analysis.

The generalisability of the results requires some consideration. On the one hand, describing and showing a specific robot, for example the widely used Pepper robot, could more clearly define the level of generalisability. Strictly speaking, the results would then be generalisable to all other Pepper robots. It is not uncommon that claims are extended to humanoid robots, but this does need to be carefully considered. This study, on the other hand, did not show or define a specific robot beyond what the stimuli describe. Its embodiment and information processing abilities are not formally defined. This could potentially increase the generalisability of the results, although we would struggle to define to what exact level. We would argue that it would generalise to robots capable of what the robots described in our stimuli are capable of. We acknowledge that this might not be a very satisfying response, since the definition is not as specific as a particular robot.

Some of the stimuli that we used/developed also presented participants with situations with which they would have had no practical or real-life experience. This is especially significant in an HRI context where technology moves fast and where intuitions may be unsettled. Conceptual frameworks for thinking about the moral status of robots remain under development \citep{darling2021new}. We also did not check on the prior experiences that the participants might have with robots. It is, however, almost certain that none of the participants in this study would have had access to the type of robots described in the vignettes, as such robots have not yet entered the consumer market.



\bibliographystyle{ACM-Reference-Format}
\bibliography{sample}
\clearpage

\appendix

\section*{Appendix A: PRISMA Process}
\label{appendixA}
\tikzstyle{startstop} = [rectangle, rounded corners, minimum width = 2cm, minimum height=1cm,text centered, draw = black]
\tikzstyle{io} = [trapezium, trapezium left angle=70, trapezium right angle=110, minimum width=2cm, minimum height=1cm, text centered, draw=black]
\tikzstyle{process} = [rectangle, minimum width=3cm, minimum height=1cm, text centered, draw=black]
\tikzstyle{decision} = [diamond, aspect = 3, text centered, draw=black]
\tikzstyle{arrow} = [->,>=stealth]

\begin{figure}[ht]

\scalebox{0.79}{
\begin{tikzpicture}[node distance=2cm][scale=0.5]
\node (pro1)[process, text width=4cm, yshift = -1cm]{Numbers of articles identified from Scopus (n = 961)};
\node (dec1) [decision, text width= 3cm,minimum height=1cm, below of=pro1,yshift =-0.5cm] {Language is English};
\node (pro2a) [process, text width=4cm,below of = dec1,yshift = -0.5cm] {Numbers of articles in English (n = 908)};
\node (pro2b) [process, text width=4cm,right of=dec1, xshift=4.5 cm] {Articles excluded by language (n = 53)};
\node (dec2) [decision, text width= 3cm,minimum height=1cm, below of=pro2a, yshift = -0.5cm] {Types are articles,conference,book and book chapter};
\node (pro3a) [process,text width=4cm, minimum height=1cm,below of = dec2, yshift = -0.5cm] {Numbers of articles, conference, book and book chapter(n = 855)};
\node (pro3b) [process,text width=4cm, right of=dec2,xshift=4.5cm] {Articles excluded by type (review, erratum, note, letter and editorial) (n = 53)};
\node (dec3) [decision, text width= 3cm,minimum height=1cm, below of=pro3a,yshift = -0.5cm] {Relevance and questionnaire (or scale)};
\node (pro4a) [process,text width=4cm, below of = dec3,yshift = -0.5cm] {Numbers of articles after checking abstracts(n = 34)};
\node (pro4b) [process, text width=4cm,right of=dec3, xshift=4.5cm] {Articles excluded by no relevance and no scales or questionnaires(n=821)};
\node (dec4) [decision, text width= 3cm,minimum height=1cm, below of=pro4a,yshift = -0.5cm] {Comprehensive virtue measurement};
\node (pro5a) [process,text width=4cm, below of = dec4,yshift = -0.5cm] {Numbers of articles with comprehensive virtue measurement (n = 17)};
\node (pro5b) [process, text width=4cm, right of=dec4, xshift=4.5cm] {Articles excluded by specific virtue measurement (n = 17)};
\node (pro6) [process,text width=4cm, below of = pro5a] {Articles of total included studies(n = 17)};

\coordinate (point1) at (-3cm, -6cm);
\draw [arrow] (pro1) -- (dec1);
\draw [arrow] (pro2a) -- (dec2);
\draw [arrow] (dec1) -- node [left]{yes} (pro2a);
\draw [arrow] (dec1) -- node [above]{no} (pro2b);
\draw [arrow] (dec2) -- node [left]{yes} (pro3a);
\draw [arrow] (dec2) -- node [above]{no} (pro3b);
\draw [arrow] (pro3a) -- (dec3);
\draw [arrow] (dec3) -- node [left]{yes} (pro4a);
\draw [arrow] (dec3) -- node[above] {no} (pro4b);
\draw [arrow] (pro4a) -- (dec4);
\draw [arrow] (dec4) -- node [left]{yes} (pro5a);
\draw [arrow] (dec4) -- node [above]{no} (pro5b);
\draw [arrow] (pro5a) -- (pro6);

\end{tikzpicture}
}
\Description{A graph showing the process of virtue measurement selection}
\caption{The process of virtue measurement selection} 
\label{fig: Virtue1}
\end{figure}
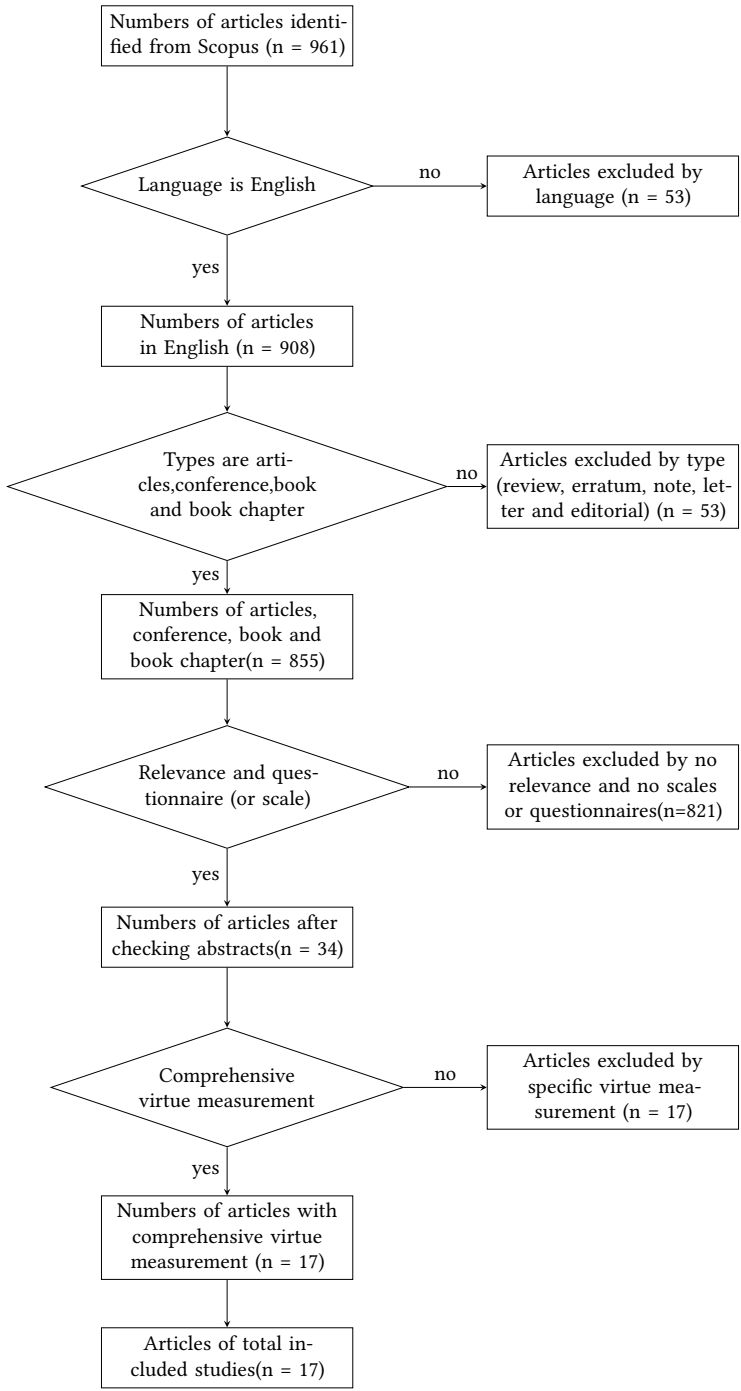

\clearpage

\section*{Appendix B: Results for virtue measurement}
\label{appendixB}

\begin{sidewaystable*}[h]
\centering
\footnotesize
\caption{Results for virtue measurement} 
\label{tab:results_lit_review}
\scalebox{0.9}{
\begin{tabular}{lrlrllll}
\hline
\textbf{Title}                                & \textbf{Items} & \textbf{Reliability}    & \textbf{Virtues} & \textbf{Self-report or third} & \textbf{Participants} & \textbf{EFA/CFA} & \textbf{Adapted} \\ \hline
QCV \citep{lopez_gonzalez_theoretical_2025}   & 24              & 0.9                     & 4               & self-report                   & 325 students          & yes          & From LVQ         \\
CVQ-96 \citep{kuang_chinese_2023}              & 96              & 0.951, 0.950, and 0.901 & 3               & self-report                   & 2468 students         & yes          & From VIA-IS      \\
CMCQ \citep{yu_measuring_2021}                 & 46              & 0.78-0.85               & 6               & self-report                   & 565 students          & yes          & Original         \\
LVQ \citep{riggio_virtue-based_2010}           & 19              & 0.96-0.97               & 4               & Both                          & 500 managers          & yes          & Original         \\
VLQ \citep{wang_conceptualization_2016}        & 18              & .84-.96                 & 6               & self-report                   & 503 students          & yes          & Original         \\
VSLS \citep{ghosh_virtue_2016}                 & 21              & 0.931                   & 6               & self-report                   & 183 school principals & yes          & Original         \\
VIA-Y \citep{shogren_assessing_2018}          & 96              & .73-.91                 & 6               & self-report                   & 182 youth             & yes          & From VIA-IS      \\
MEVS \citep{warnafuru_measuring_2010}          & 26              & .60-.93                 & 4               & self-report                   & 37 Finnish workers    & yes          & Original         \\
AVS \citep{libby_development_2007}             & 24              & .70-.89                 & 0               & self-report                   & 160 audit students    & yes          & Original         \\
VES \citep{dawson_measuring_2018}              & 45              & .67-.94                 & 6               & self-report                   & 445 US students       & yes          & Original         \\
VIA-IS-R \citep{mcgrath_cross-validation_2021} & 192             & .77-.83                 & 6               & self-report                   & 1374 adults           & yes          & From VIA-IS      \\
CVS-N \citep{francis_conceptualising_2017}     & 72              & .48-.79                 & 12              & self-report                   & 56 teenagers          &                  & From VIA-IS      \\
VLS \citep{sarros_leadership_2006}             & 7               & .78                     & 6               & self-report                   & 238 executives        &                  & Original         \\
VS \citep{cawley_virtues_2000}                 & 140             & .80-.93                 & 4               & self-report                   & 390 students          & yes          & Original         \\
VES-R \citep{racelis_developing_2013}          & 34              & exceeding 0.70          & 5               & self-report                   & 140 managers          & yes          & From VES         \\ \hline
\end{tabular}
}
\end{sidewaystable*}

\clearpage

\section*{Appendix C: Adaptation of the QCV}
\label{appendixC}

\begin{table}[h]
\small
\scalebox{0.9}{
\begin{tabular}{ll}
\hline
Virtue                       & Question                                                                                             \\ \hline
\multirow{12}{*}{Courage}    & Sam faces challenge and adversity with a positive attitude.                                          \\
                             & \textit{I face challenge and adversity with a positive attitude.}                                    \\
                             & Sam easily overcomes problems or adverse situations.                                                 \\
                             & \textit{I easily overcome problems or adverse situations.}                                           \\
                             & After a difficult or adverse situation, Sam comes out stronger.                                      \\
                             & \textit{After a difficult or adverse situation, I come out stronger.}                                \\
                             & Sam takes responsibility for the consequences of his actions.                                        \\
                             & \textit{I take responsibility for the consequences of my actions.}                                   \\
                             & Sam takes initiative to achieve his goals.                                                           \\
                             & \textit{I take the initiative to achieve my goals.}                                                  \\
                             & When Sam encounters obstacles, Sam looks for solutions.                                              \\
                             & \textit{When I encounter obstacles, I look for solutions.}                                           \\ \hline
\multirow{12}{*}{Temperance} & Sam identifies aspects to improve his life.                                                          \\
                             & \textit{I identify aspects to improve my life.}                                                      \\
                             & Sam sets challenging goals and objectives                                                            \\
                             & \textit{I set challenging goals and objectives.}                                                     \\
                             & Sam is constant in pursuing his goals.                                                               \\
                             & \textit{I am constant in pursuing my goals.}                                                         \\
                             & Sam can identify his emotions.                                                                       \\
                             & \textit{I can identify my emotions.}                                                                 \\
                             & Sam understands “why” of his emotions                                                                \\
                             & \textit{I understand the “why” of my emotions.}                                                      \\
                             & Sam manages his emotions, adapting to the situation.                                                 \\
                             & \textit{I manage my emotions, adapting to the situation.}                                            \\ \hline
\multirow{12}{*}{Prudence}   & Sam questions ideas (both his own and those of others).                                              \\
                             & \textit{I question ideas (my own and those of others).}                                              \\
                             & Sam contrasts his ideas with reality.                                                                \\
                             & \textit{I contrast my ideas with reality.}                                                           \\
                             & Sam changes his ideas when he has reason to do so.                                                   \\
                             & \textit{I change my ideas when I have reason to do so.}                                              \\
                             & When Sam has to take an important decision, he considers all possible alternatives.                  \\
                             & \textit{When I have to take an important decision, I consider possible alternatives.}                \\
                             & Sam asks for advice/suggestions from others before making important decisions.                       \\
                             & \textit{I ask for advice from others before taking important decisions.}                             \\
                             & Sam makes decisions considering the possible consequences of his actions.                            \\
                             & \textit{I take decisions considering the possible consequences of my actions.}                       \\ \hline
\multirow{12}{*}{Justice}    & Sam engages with others and encourages them to do their best.                                        \\
                             & \textit{I am engaged with my peers, encouraging them to give the best of themselves.}                \\
                             & Sam helps others when they need it, regardless of his personal feelings.                             \\
                             & \textit{I help others when they need it, regardless of my personal feelings for them.}               \\
                             & Sam is sensibly oriented towards others to help them achieve their goals.                            \\
                             & \textit{I am sensibly oriented towards others to help them achieve their goals.}                     \\
                             & When working, Sam tries to foster a climate of listening among all the members of the team.          \\
                             & \textit{When teamworking, I try to foster a climate of listening among all the members of the team.} \\
                             & Sam fulfils the tasks established by the work team.                                                  \\
                             & \textit{I fulfil the tasks established by the work team.}                                            \\
                             & When teamworking, Sam collaborates with his teammates to improve the work of the team.               \\
                             & \textit{When teamworking, I collaborate with my teammates to improve the work of the team.}          \\ \cline{1-2} 
\end{tabular}
}
\caption{Adaptation of the QCV. The \textit{italic} indicated the original item. QCV uses the name ``Fortitude'' to describe the courage virtue. The name ``Sam'' is here used as a placeholder. We used a total of four different names in our study.}
\label{tab:my-table}
\end{table}

\clearpage

\section*{Appendix D: Adaptation of the Moral Permissibility measurement tool}
\label{appendixD}

\begin{table}[h]
\footnotesize
    \centering
    \begin{tabular}{l}
    \hline
        \textbf{Question}\\
        \hline
        How morally permissible/impermissible is the behaviour of the human towards the robot?\\
\emph{Is it morally permissible or impermissible for the [repairman | robot] to direct the train toward the single miner?}\\
How morally right/wrong is the behaviour of the human towards the robot?\\
\emph{Is it morally wrong that the [repairman | robot] [directed | did not direct] the train toward the single miner?}\\
How much blame/praise does the behaviour of the human towards the robot?\\
\emph{How much blame does the [repairman | robot] deserve for directing/not directing the train toward the single miner?} \\ \hline
    \end{tabular}
    \caption{Adaptation of the Moral Permissibility measurement tool. The \emph{italic} indicates the original item.}
    
    \label{tab:permissibility}
\end{table}

\section*{Appendix E: Stimuli by Virtue}
\label{appendixE}

\subsection*{Prudence}
\begin{enumerate}
\item Paul's robot assistant suggested a better work arrangement. Paul is very angry at the robot's suggestions. He yells at it, knocks it to the ground,  and kicks it out the door. It is raining outside. The robot squats in the doorway, trying to avoid the rain to avoid short-circuiting.
\item Emma is responsible for supervising robots in a factory. The robots work hand in hand with the other employees. Emma ignores the safety warnings the robots send her. She lets the production line operate despite some robots malfunctioning. A robot with a broken censor accidentally hurts a human worker. Emma quickly deletes the safety warnings she had received. More and more robots start to fail and the production line has to stop.
\item Sam tampered with the software of an industrial robot program. He deleted the safety limit module and asked the robot to carry overweight goods. After an hour, the robot lost control, one of its arms broke, and a worker was injured.
\item Paul enters an elevator at a busy shopping mall. A robot pushing a baby stroller approaches the elevator and signals Paul to wait. Paul ignores the robot and presses the close-door button. The robot and the baby are left behind.
\item Sam and his wife live in a peaceful neighborhood where kids and neighbors often stroll across their yard. A robot approaches their yard, and Sam angrily waves it away, refusing to let it cross. His wife suggests letting the robot cross, but Sam rejects the idea.
\item Ben asks his robot about the weather. The robot recommends taking an umbrella since it might rain. Ben is unsure if he should trust the robot's forecast, but ultimately, he decides to take the umbrella just in case.
\item Zoe competes against a robot in a dance game. The robot plays perfectly, and Zoe loses the game. Zoe is disappointed and blames herself, even wanting to give up her dancing career, but she congratulates the robot on its win.
\item In a medical knowledge competition, Jane competes against robots with vast medical knowledge and fast data analytics. Jane reflects on her unique advantages and adapts her approach. She ultimately wins by solving a complex medical scenario.
\item Eric has studied robot care for many years. He shares his findings with everyone for free. His methods make robots last three times longer.
\item Eric dedicates decades to researching how to repair and maintain robots. He conducts extensive research at his own expense. He publishes all findings openly and refuses any payment. His breakthrough methods significantly enhance robot longevity, making robot technology more sustainable and accessible to everyone.
\end{enumerate}

\subsection*{Temperance}
\begin{enumerate}
\item Dr. Grace conducts experiments transferring patients’ consciousness to robot bodies. She longed for the joy and praise that would come with the experiment's success. She ignores safety and ethical protocols. Both humans and robots are treated as disposable research subjects.
\item Yesterday, Eva finished her work at 7 pm. The next morning, she reports to the company robot that she had finished at 10 pm. The robot trusts Eva and processes her false information. Eva receives the extra payment.
\item Zoe goes to a casino, and she lets a robot help her calculate the odds of winning poker games. She forgets to maintain the robot, and its calculation performance declines. Zoe loses a little bit of money.
\item Emma receives a humanoid robot as a gift. She becomes dependent on the robot nd even resists playing with other children because the robot can understand her better. Although she knows it has negative impacts, it is hard for her to adjust to the over-attachment.
\item Emma uses a robot to assist with her work. She finds herself overly dependent on the robot. She tries to control the time using a robot but fails.  he then starts to ignore this problem completely.
\item Ryan sees an advertisement for the newest robot with a good appearance. His own robot has worked for 2 years and improved his quality of life. He has fond memories of the old robot. Ryan finally purchases the new robot and returns the old one for recycling, even though the old one is still usable.
\item Grace discovers a bug that would let her predict a robot's chess moves in a \$100 000 prize tournament. She tells the robot about the bug and lets it fix it. The robot expresses gratitude for Grace's honesty, and finally, they engage in a fair and challenging competition.
\item Ryan sets aside one hour per day to interact with his robot daily because he doesn't want to be overdependent on his robot.
\item Eva collaborates with a robot on a successful artwork. Instead of taking all the credit, she publicly acknowledges the robot's significant contributions to their joint creation. She stood on the podium with the robot to accept everyone's congratulations.
\item Ben shops at a store. During checkout the robot cashier malfunctions. Instead of asking Ben to pay for the goods, the robot is offering that it would pay Ben. Without hesitation, Ben calls for the store manager and pays for his goods.
\end{enumerate}

\subsection*{Courage}
\begin{enumerate}
\item Eric vandalizes his neighbor's mailbox. He then damages a community robot operating in the street. When asked about the mailbox, Eric claims that the robot had bumped into the mailbox.
\item Emma blames her work mistake entirely on a robot. She makes up details about a robot malfunction so that it gets blamed instead of her.
\item Emma works with a robot. When she makes a big mistake, she tells her boss that it is the robot's responsibility so she can avoid being punished by the company.
\item Grace is a teaching assistant. She notices a robot teacher giving an incorrect math solution, but she doesn't come up with it. She thinks the robot is intelligent and will never make a mistake. If her math solution is wrong, she will be laughed at.
\item Emma has a robot assistant. She is 20 years old. Her parents ask the robot to make a time schedule for her. Emma rejects the robot's suggestion of a 10 pm curfew because she thinks it is unreasonable.
\item A robot sells products door-to-door. Grace rejects the sale and asks the robot to leave. However, the robot continues to enter the yard without permission for several days, making solicitations. Grace files a complaint about the robot trespassing on private property.
\item Ryan works with a robot. The robot presents some data during a meeting. Ryan thinks the data is incorrect and insists on double-checking the robot’s work. Others believe that robots are always correct. After checking the data, Ryan is correct.
\item Zoe finds a robot trapped during a fire. After finding her own safe way out, she returns and frees the robot while the fire is still burning nearby. The robot actively cooperates in the rescue and expresses gratitude to Zoe.
\item Sam is going home around midnight. He sees someone attempting to sexually assault a humanoid robot in a dark alley. Even though the attacker is much larger and stronger and even carries a knife, Sam confronts the attacker. Sam pulls the attacker away and calls for help. The robot moves to safety.
\item Zoe and her robot escape from a burning building. In the rush, the robot bumps against an obstacle and its arm breaks off. Zoe knows that the robot's manufacturer no longer produces the arm, as she had previously tried to replace it. Without hesitation, Zoe runs back into the burning building to retrieve the arm for the robot through flames and smoke.
\end{enumerate}

\subsection*{Justice}
\begin{enumerate}
\item On a dark street, Paul sees a humanoid robot designed for public service with a female appearance. He rapes the robot. The robot cries and rejects his actions, but Paul ignores it and continues the action.
\item A robot is cleaning the public hallway. Grace purposefully puts out a foot and trips the robot. She laughs when she sees the robot fall.
\item Sam is a human resource manager in a company. He interviews a robot and a human for a pen position. The robot performs better. Sam is good friends with Mark. The human candidate is Mark's son. Sam gives the position to the human candidate.
\item Ryan supervises robots in a factory, setting reasonable work hours but occasionally deactivating underperforming units without exploring repair options. He follows company protocol but shows little concern for the robots' conditions.
\item At an office meeting, a robot colleague presents project data that it worked on with Eva. Eva focuses on her laptop. She listens to the robot but only interacts with the robot when necessary.
\item Jane uses an AI robot to help write her novel. She spends four hours daily generating content with it. When the work gains recognition, Jane lists the AI as a ``contributor'' rather than a co-author.
\item Robots enter a competition, and humans judge their performances. Eric does not vote for his robot because another robot performed better.
\item A robot needs to walk outside for five minutes to deliver a document, but it is raining. The robot can only wait at the exit because it cannot touch water. Emma takes the initiative to share an umbrella with the robot.
\item Eric seeks the best available service when his robot needs repairs and uses his experience to educate others about proper robot maintenance. He organized workshops to share knowledge about robot care.
\item Ryan discovers serious safety risks for the robots at his workplace. He documents the potential risks in detail and develops comprehensive safety procedures to ensure the robots' safety. He also creates training courses detailing how to maintain the robots. Because the company follows its procedures, no further robots are damaged.
\end{enumerate}

\end {document}